\documentclass[prb,reprint,superscriptaddress,twocolumn,showpacs,floatfix]{revtex4-1}
\usepackage{graphicx,color}
\usepackage{amsmath,amssymb,bm}
\usepackage{epstopdf}
\setcitestyle{square,numbers}

\newcommand{\mc}{\mathcal}
\newcommand{\nd}{\vphantom{\dag}}

\usepackage[plainpages=false,pdfpagelabels,colorlinks=true,linkcolor=red,urlcolor=blue,citecolor=blue,pdftitle={Title},pdfauthor={},pdfdisplaydoctitle=true,pdfduplex=DuplexFlipLongEdge]{hyperref}

\begin{document}
\title{Matrix-product-state method with a dynamical local basis optimization for bosonic systems out of equilibrium}

\author{C. Brockt}
\email[E-mail: ]{christoph.brockt@itp.uni-hannover.de}
\affiliation{Institut f\"{u}r Theoretische Physik, Leibniz Universit\"{a}t Hannover, Appelstrasse 2, D-30167 
Hannover, Germany}
\author{F. Dorfner}
\affiliation{Department of Physics and Arnold Sommerfeld Center for Theoretical Physics,
Ludwig-Maximilians-Universit\"at M\"unchen, D-80333 M\"unchen, Germany}
\author{L. Vidmar}
\affiliation{Department of Physics and Arnold Sommerfeld Center for Theoretical Physics,
Ludwig-Maximilians-Universit\"at M\"unchen, D-80333 M\"unchen, Germany}
\author{F. Heidrich-Meisner}
\affiliation{Department of Physics and Arnold Sommerfeld Center for Theoretical Physics,
Ludwig-Maximilians-Universit\"at M\"unchen, D-80333 M\"unchen, Germany}
\author{E. Jeckelmann}
\affiliation{Institut f\"{u}r Theoretische Physik, Leibniz Universit\"{a}t Hannover, Appelstrasse 2, D-30167 
Hannover, Germany}

\begin{abstract}
We present a method for simulating the time evolution of one-dimensional
correlated electron-phonon systems which combines the time-evolving block
decimation algorithm with a dynamical optimization of the local basis.
This approach can reduce the computational cost by orders of magnitude when
boson fluctuations are large.
The method is demonstrated on the nonequilibrium Holstein polaron by
comparison with exact simulations in a limited functional space and on
the scattering of an electronic wave packet by local phonon modes.
Our study of the scattering problem reveals a rich physics including
transient self-trapping and dissipation.

\end{abstract}

\date{\today}


\maketitle

Phonon degrees of freedom play an important role in the nonequilibrium properties of correlated materials.
In particular, time-resolved spectroscopy~\cite{baso11,orenstein12}, photoinduced phase transitions~\cite{nasu04,yone08},
and transport through low-dimensional or molecular junctions~\cite{galp07,osor08,zimb11}
call for theoretical investigations of the nonequilibrium dynamics of charge carriers coupled
to lattice vibrations.
Quantum lattice models such as the Holstein model~\cite{hols59} are often used to describe
the low-energy properties of strongly correlated electron-phonon (EP) systems.
Analytical studies of these systems out of equilibrium are very difficult and reliable results are scarce.
Therefore, theorists often turn to numerical methods to investigate them~\cite{jeck07,assa08,hohe08}.
However, accurate numerical simulations of correlated lattice systems are very challenging 
because of the rapid increase of the Hilbert space dimension with system size and phonon number
fluctuations.
Similarly, computing the nonequilibrium dynamics of correlated bosons is a 
significant challenge in a great variety of physical systems such as nonlinear optical systems~\cite{gruj12,caru13},
quantum dissipative systems~\cite{weiss2008,yao13}, and low-energy models of quantum chromodynamics~\cite{berg14}.

In this paper we present a method for simulating the time evolution of one-dimensional (1D)
lattice models with strongly fluctuating bosonic degrees of freedom for long periods of time.
It combines the time-evolving block decimation (TEBD)~\cite{vida03,vida04} with
a local basis optimization (LBO)
approach~\cite{zhan98} to reduce the computational cost significantly.
The key idea is to optimize the local bases for the bosonic degrees of freedom dynamically and adaptively.
The accuracy of the method is first demonstrated by comparison with reliable results for a nonequilibrium 
polaron problem~\cite{dorf15}. Then its performance is illustrated with a study of wave packet scattering by a 
small EP-coupled structure.

In quantum lattice models phonon degrees of freedom are represented by bosonic sites.
As the Hilbert space of a single bosonic site is already infinite, it must be truncated to a subspace
of dimension $M$ from the start in wave-function-based numerical approaches~\cite{jeck07}. The most common choice is to use the lowest $M$ eigenstates of a
(well chosen) boson number operator $b^\dagger b$
defining a bare
boson basis. 
Then exact diagonalization or exact time propagation can be easily performed but the computational
cost increases very rapidly with $M$ and exponentially with the number of lattice sites. 
Matrix-product-state (MPS) algorithms, such as the density-matrix renormalization group
(DMRG)~\cite{whit92b,whit93a,scho05,jeck08a,scho11} and TEBD, allow us to treat 1D systems at a 
lower computational cost and thus to investigate much larger 
systems. However, the computational effort still increases as $M^3$.
Therefore, most applications have been restricted
to problems with small phonon fluctuations ($M \alt 10$), in particular for nonequilibrium problems~\cite{mats12}.

Instead of a bare boson basis of dimension $M$, one can describe a quantum state $\vert \psi \rangle$ using an optimal local basis of 
dimension $M_{O} \leq M$, which is defined as the eigenbasis of the reduced density matrix of $\vert \psi \rangle$ 
for the bosonic site~\cite{zhan98}.
This approach is very efficient for ground-state calculations because a sufficient accuracy can be reached with
a small optimal basis even when a very large bare basis would be required. As the optimal basis must be calculated self-consistently,
the total computational cost still rises with $M$ but only linearly. Thus ground-state calculations can be carried out
with exact diagonalization or MPS algorithms for systems with $M \geq 10^3$ using only moderate computer 
resources~\cite{zhan98,barf02,jeck07,guo12}.

The LBO has never been combined successfully  with MPS methods to study EP systems
out of equilibrium but for
a very recent study of the spin-boson model~\cite{schr15}.
The key problem is that the local optimal basis depends on the represented quantum state $\vert \psi (t) \rangle$ and thus evolves
with time.  This is clearly seen in our recent study of the optimal boson basis for a nonequilibrium polaron problem
(see Figs.~18-20 in Ref.~\cite{dorf15}).
Therefore, we have developed an algorithm which allows us to optimize the local basis dynamically for the evolving target state 
$\vert \psi (t) \rangle$.

\begin{figure}[tb]
\includegraphics[width=1.00\columnwidth]{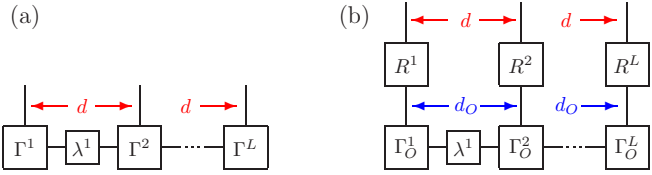}
\caption{(Color online)
Graphical representation of the MPS (a) in the original TEBD algorithm and (b) in the TEBD-LBO algorithm.}
\label{fig:sketch}
\end{figure}

We have implemented this approach within the TEBD algorithm~\cite{vida03,vida04}  which is one
of the simplest time-dependent MPS methods~\cite{whit04,dale04,schm04a}.
For a chain with $L$ sites the MPS representation of a quantum state $\vert \psi \rangle$ in an occupation number basis is
\begin{equation}
\label{eq:mps}
\psi(k_1,\dots, k_{L}) = \Gamma^{1,k_1} \lambda^{1} \Gamma^{2,k_2}
\lambda^{2} \dots 
\lambda^{L-1}  \Gamma^{L,k_{L}}  ,
\end{equation}
where the indices $k_j$ label the basis states of the $d_j$--dimensional Hilbert space representing 
the degrees of freedom on the lattice site $j \in \{1,\dots,L\}$. 
(For a bosonic site, $d_j=M$.)
The entanglement between two parts of the lattice (e.g., the sites $\{1,...,j\}$ and the sites $\{j+1,...,L\}$)
is encoded in the $D_j$--dimensional positive definite diagonal matrices $\lambda^j$. Hence the matrices $\Gamma^{j,k_{j}}$
have dimensions $D_{j-1} \times D_{j}$ (with $D_0=D_L=1$).
This MPS is represented graphically in Fig.~\ref{fig:sketch}(a). 
We call $D = \max\ \{D_1,...,D_{L-1}\}$  the bond dimension of the MPS and  $d =  \max\ \{d_1,...,d_{L}\}$ its local dimension.

Using orthogonality relations for the matrices $\Gamma$ and $\lambda$,
the matrix elements of the reduced density matrix $\rho^{j}$ for the site $j$ are given by
\begin{equation}
\label{eq:densmat}
\rho^{j}_{k_j,k'_j}  =   {\text{Tr}} \left [ \left ( \Gamma^{j,k'_j} \right )^{\dagger} 
\left ( \lambda^{j-1} \right )^2  \Gamma^{j,k_j} \left ( \lambda^{j} \right )^2 \right ] .
\end{equation}
The eigenbasis of this $d_j \times d_j$ matrix is called the optimal local basis. 
The unitary transformation from the optimal to the bare basis representation is denoted by $R^j$
\begin{equation}
\label{eq:transf}
\Gamma^{j,k_j} =  \sum_{s_j=1}^{d_O}  R^{j}_{k_j,s_j} \Gamma_O^{j,s_j}. 
\end{equation}
This transformation is exact if  $d_O=d_j$.
The matrices $\lambda^{j}$ are not affected by the basis change.
The new MPS structure is illustrated in Fig.~\ref{fig:sketch}(b).
This  transformation of the matrices $\Gamma$ in the TEBD algorithm
is similar to the approach proposed for a variational MPS~\cite{guo12,bruo14}.
Each optimal basis state has a weight (eigenvalue) in the interval $[0,1]$.
We can thus approximate the original state~(\ref{eq:mps}) using only the $d_O (\leq d)$
eigenstates with the highest weights.

For a Hamiltonian which includes only on-site and nearest-neighbor interactions $H = \sum_j H_{j,j+1}$,
the time-evolution of the MPS~(\ref{eq:mps}) can be decomposed into successive local updates with a time step $\tau$
using a Trotter-Suzuki decomposition (TSD) of the time evolution operator:
$\tilde{\psi} =  e^{-i\tau H_{j,j+1}} \psi$, where the local operator $H_{j,j+1}$ acts only on a single bond (i.e., sites
$j$ and $j+1$)~\cite{vida03,vida04}.
Each local update is a unitary transformation which modifies the two matrices $\Gamma$ and the one matrix $\lambda$ associated with 
a bond.
In practice, we use a second-order TSD
resulting in an error $\mathcal{O}(\tau^3)$ per time step or $\mathcal{O}(\tau^2)$ for a finite period of time.

Here we only explain how we perform the local update in our TEBD-LBO algorithm
as our method is otherwise  identical to the original TEBD~\cite{vida03,vida04}.
We assume that we know the MPS~(\ref{eq:mps}) at a given time $t$ in its truncated optimal local basis, 
i.e., we know the $\Gamma_O$, $\lambda$ and $R$ matrices
with $d_j \leq d_O \leq d$. A local update for a single bond consists of four steps.
First, we build the rank-four tensor 
  \begin{equation}
    \label{phi}
    \phi_{\alpha_1,\alpha_2}^{k_j,k_{j+1}} = [\lambda^{j-1} \Gamma^{j,k_j} \lambda^{j} 
    \Gamma^{j+1,k_{j+1}} \lambda^{j+1}]_{\alpha_1,\alpha_2} ,
  \end{equation}
  where the $\Gamma$ matrices are given by~(\ref{eq:transf}) and the indices $\alpha_{1,2}=1,\dots,D$ number
  the matrix rows and columns.
Then, in a second step we carry out the time evolution as done in the original TEBD algorithm,
  \begin{equation}
  \label{phiprime}
	\tilde{\phi}_{\alpha_1,\alpha_2}^{{l}_j,{l}_{j+1}} 
	= \sum_{k_j,k_{j+1}} U_{k_j,k_{j+1}}^{{l}_j,{l}_{j+1}}\ \phi_{\alpha_1,\alpha_2}^{k_j,k_{j+1}},
 \end{equation}
where $U$ denotes the $d_jd_{j+1} \times d_jd_{j+1}$ matrix representation of the local time-evolution operator
$e^{-i\tau H_{j,j+1}}$ in the bare basis.
Generally, the computational cost for this step is $\mc{O}(d^4 D^2)$ but
it can be reduced to $\mc{O}(d^2 D^2)$ using the sparseness of the matrix representation of~(\ref{hamiltonian})
in a bare boson basis.
In the third step, we compute the local reduced density matrix~(\ref{eq:densmat}) from the tensor $\tilde{\phi}$ using the relation
\begin{equation}
\rho^{j}_{k_j,k'_j}  =  \sum_{k_{j+1},\alpha_1,\alpha_2}  {\text{Tr}} \left [ \tilde{\phi}_{\alpha_1,\alpha_2}^{{k}_j,{k}_{j+1}} 
\left ( \tilde{\phi}_{\alpha_1,\alpha_2}^{{k'}_j,{k}_{j+1}} \right )^{*} \right ]
\end{equation} 
and then diagonalize it.
This yields the new optimal bases for the
sites $j$ and $j+1$, i.e., new transformations $\tilde{R}^j$ and $\tilde{R}^{j+1}$.
We discard the optimal eigenstates with a negligible weight  (e.g., lower than $10^{-13}$)
and thus obtain a new truncated optimal basis of dimension $\tilde{d}_O \leq d$. 
The tensor $\tilde{\phi}$ is then projected into the new optimal basis.
Finally, in a fourth step the new matrices $\tilde{\Gamma}_O^j$, $\tilde{\Gamma}_O^{j+1}$, and $\tilde{\lambda}^j$ 
are calculated from the projected tensor $\tilde{\phi}_O$ exactly as in the original TEBD algorithm. 
Hence we obtain the MPS representation of the state $\tilde{\psi}$ at time $t+\tau$
in its optimal local basis.

In summary, we repeatedly propagate the wave function~(\ref{eq:mps})  in a bare local basis to enlarge the effective Hilbert space
and then project it onto a new effective Hilbert space using the LBO to control the dimension of the MPS.
If $d_O \alt d$, the total computational effort scales as $d^3 D^3$ exactly as with a bare basis.
If a small optimal basis is sufficient ($d_O \ll d$), however, our algorithm scales as the largest
of $d_O^3 D^3$ (the computational cost of TEBD in the optimal basis) and $d^3 D^2$ (the computational cost for adapting
the optimal basis)
and thus it is significantly faster than a bare basis simulation.
Contrary to the linear scaling of ground-state methods with $d$~\cite{zhan98,jeck07,guo12}, however, the computational cost 
still increases as $d^3$ but the prefactor is reduced significantly, in particular by a factor $\propto 1/D$. 
Therefore, the advantage of the TEBD-LBO will become
more pronounced for problems with a large block entanglement, i.e., $D \gg 1$.

Next we turn to two applications of our TEBD-LBO method to EP systems out of equilibrium.
We consider an $L_H$-site Holstein chain~\cite{hols59} connected  at each end to tight-binding leads with $L_{\text{TB}}$ fermion sites.
The Hamiltonian of the full system (with $L=L_H +2 L_{\text{TB}}$ sites) is 
\begin{eqnarray}
 \label{hamiltonian}
 H & = & -t_0  \sum_{j=1}^{L-1} \left( c_{j}^{\dag} c_{j+1}^{\nd} + c_{j+1}^{\dag} c_{j}^{\nd} \right) \\
 & & + \sum_{j=L_{\text{TB}}+1}^{L_{\text{TB}}+L_H}
 \left [ \omega_0 \, b_{j}^{\dag}  b_{j}^{\nd} - \gamma \left( b_{j}^{\dag} + b_{j}^{\nd} \right) n_{j}^{\nd} \right ], \nonumber 
\end{eqnarray}
where $b_{j}$ and $c_{j}$ annihilate a phonon (boson) and a (spinless) fermion on site $j$, respectively,
 and $n_{j}^{\nd} = c_{j}^{\dag} c_{j}^{\nd}$. Thus $d=2M$ in this model. 
The model parameters are the phonon frequency $\omega_0 > 0$, the EP coupling 
$\gamma$ and the hopping integral $t_0$. 
We work with $\hbar=1$ and set the energy scale $t_0=1$, thus the time unit is $\hbar/t_0=1$.

Here we restrict ourselves to the nonequilibrium dynamics of an electron coupled to phonons (i.e., the polaron dynamics), 
which has recently become a widely studied topic~~\cite{ku07,kenn10,luo10,fehs11,vidm11,golez12b,li13,dey14,wern15,dorf15,sayy15,mish15,zhou15}.
One-electron problems have MPS with low bond dimensions $D$,
which can easily be simulated on a workstation
when the effective local dimension is small ($d$ or $d_O \alt 10$).
Thus they provide us with a practical test field for our TEBD-LBO method.
Typically, we use $D \leq 30$ or kept all block eigenstates with weight $> 10^{-15}$
in combination with a time step $\tau$ as small as $10^{-3}$
to keep TEBD errors (induced by the TSD and the truncation of the bond dimensions) under control.
The conservation of the electron number is used to decompose the matrices $\Gamma, \lambda, \rho$, and $\phi$ 
into block submatrices and thus speed up the calculations. Therefore, we also obtain different optimal boson states as a function
of the electronic occupation of a site.

The initial wave function contains no phonon
\begin{equation}
\vert \psi(t=0) \rangle = \sum_{j=1}^{L} f(j) \; c_{j}^{\dag} \, \vert\emptyset\rangle,
\end{equation}
with the vacuum state $\vert\emptyset\rangle$.
Thus it is only slightly entangled, as  $D=d=2$.
Naturally, these dimensions can increase significantly when the wave function evolves with time~\cite{scho11}.
Consequently, we always start our simulations with a small bare basis dimension $d$.
After every time step $\tau$, $d$ is increased if the occupation of the highest phonon state exceeds some threshold (e.g., $10^{-7}$).

\begin{figure}[t]
\includegraphics[width=1.00\columnwidth]{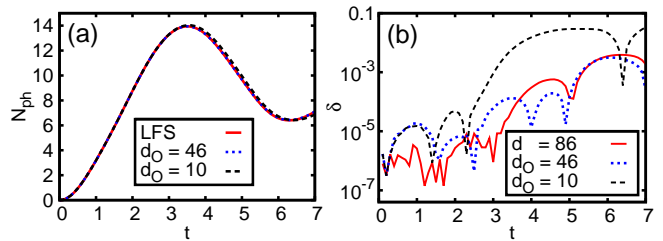}
\caption{(Color online) 
Comparison of LFS and TEBD-LBO for various $d_O$. 
(a) Time evolution of the phonon number calculated for $L=L_H=6$, $\gamma=2$ and $\omega_0=1$.
(b) Relative deviations of the TEBD-LBO data from the LFS results.
Deviations for TEBD with a bare basis of dimension $d=86$ are also shown.}
\label{fig:comparison}
\end{figure}

First, we test our algorithm on the dynamics of a highly excited electron coupled to phonons~\cite{dorf15}.
In that case, no tight-binding lead is attached to the Holstein chain ($L=L_H$) and the electron is
initially in a state $f(j) = \sqrt{2/(L+1)} \sin(K j)$ with, e.g., $K=\pi L/(L+1)$.
We have recently investigated this problem~\cite{dorf15} using very accurate simulations within a limited functional space 
(LFS)~\cite{bonc99,vidm11}
and we use these results to check the method presented here.
As an illustration, Fig.~\ref{fig:comparison} compares the time evolution of the phonon number
$N_{\text{ph}} = \langle \sum_j b_{j}^{\dag}  b_{j}^{\nd} \rangle$ calculated using TEBD-LBO and LFS.
In Fig.~\ref{fig:comparison}(a) we see that the agreement is very good  while
in Fig.~\ref{fig:comparison}(b) we observe that relative deviations become 
overall smaller when the dimension $d_O$ is increased and approach the values obtained using TEBD with a bare basis.
Therefore, the global exponential increase of deviations with time is not due to the LBO
but to intrinsic numerical errors of the TEBD and LFS methods.
However, our tests also confirm that for this problem $d_O$ must be a substantial fraction of 
the bare basis dimension $d$ (e.g.,  $d_O \approx d/4$) to achieve a similar accuracy.
Consequently, the dynamical LBO does not reduce
the computational cost significantly in comparison to the bare basis approach for this type of problem.
This is due to the relatively broad distribution of the local density-matrix eigenvalues
that we found in our previous work~\cite{dorf15}.

Second, we apply our method to the scattering of an electronic wave packet
by a small EP-coupled structure.
In that case the tight-binding leads are much longer than the Holstein chain
(we use $L_{\text{TB}}$ up to 280 and $L_H \leq 6$ sites).
The initial state is a Gaussian wave packet centered around a site $j_0$ 
in the left lead and with a positive velocity $v_0\approx 2t_0 \sin(K)$
\begin{equation}
f(j) = C \exp \left [  - \frac{(j-j_0)^2}{4 \sigma^2} \right ] \exp[i K j],
\end{equation}
where $C$ is a normalization constant, $L_{\text{TB}}-j_0 \gg \sigma \gg 1$, and $\pi > K > 0$.
For the calculations presented here, we use $\sigma = 5$ and $K=\pi/2$.
After a time $t\approx (L_{\text{TB}}-j_0) / v$ the wave packet reaches the Holstein chain where it becomes temporarily
self-trapped, and finally it is partially transmitted and reflected~\cite{suppmat}.

For this problem we find that the dynamical LBO reduces the computational
effort substantially when bosonic fluctuations are large.
For cases which can be simulated with both a bare basis and an optimized one, we already observe
speed-up factors larger than 10. For instance, for $\omega_0=0.6$ and $\gamma=2$ calculations
with a bare basis of dimension $d=124 \ (M=62)$ take $14$ times longer than with an optimized basis
of dimension $d_O=9$, but both approaches yield similar results with relative deviations smaller than $10^{-3}$.
For larger phonon fluctuations, we can only complete simulations using TEBD-LBO.
For instance, in the strong-coupling adiabatic regime ($\omega_0=0.2$ and $\gamma=2$), the required bare basis dimension
is of the order of $10^3$ but we can perform the TEBD-LBO simulation using only up to $d_O=23$ optimal states.
Therefore, the dynamical LBO allows us to study regimes that we could not treat with the standard TEBD algorithm
on our workstation.
(For comparison, dimensions $d=30$ and $D=5$ were reported in Ref.~\cite{schr15}.)

\begin{figure}[t]
\includegraphics[width=1.0\columnwidth]{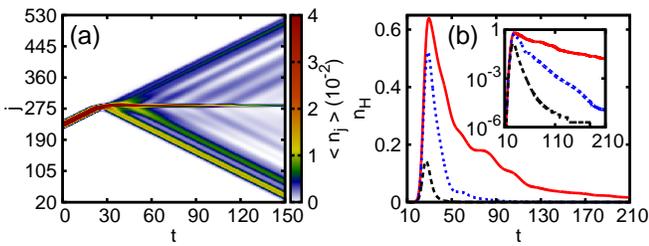}
\caption[short]{(Color online) (a) Electronic density $\langle n_j \rangle$ as a function of site $j$ and time $t$
for $L_H=6$, $\gamma=1$, and $\omega_0=2.25$. The Holstein chain corresponds to sites $j=280,\dots,285$.
(b) Total electronic density $n_H$ in a Holstein chain  of length $L_H= 6$ (red solid line), $3$ (blue dotted line) and 
$1$ (black dashed line) as a function of time $t$. 
The inset shows the same data on a logarithmic scale.
}
\label{fig:color_map_LW6}
\end{figure}

The direct injection of an electronic wave packet into an EP-coupled chain
was studied previously~\cite{ku07,fehs11} but the scenario considered here
has not been studied yet.
Thus we briefly discuss two interesting phenomena that we have observed but postpone a more thorough 
discussion to a future work.
The first phenomenon is the temporary self-trapping of the electron in the EP-coupled structure.
In Fig.~\ref{fig:color_map_LW6}(a) we see that the electron reaches the EP-coupled structure at $t\approx 30$
but that a finite density remains in that region even for $t = 150$~\cite{suppmat}. 
At several times, fractions of the wave packet leave the Holstein chain
and start to propagate in the leads.
The probability of finding the electron in the Holstein chain $n_H = \sum_{j=L_{\text{TB}}+1}^{L_{\text{TB}} + L_H} \langle n_{j} \rangle$
is shown in Fig.~\ref{fig:color_map_LW6}(b). It increases rapidly when the wave packet reaches the left edge site of the Holstein chain
at $t\approx 30$ and then decays exponentially fast for longer times. The decay rate is longer for longer chains.
Therefore, (a fraction of) the electronic wave packet becomes temporarily self-trapped in the EP-coupled structure and
is belatedly transmitted or reflected.

\begin{figure}[tb]
\includegraphics[width=1.0\columnwidth]{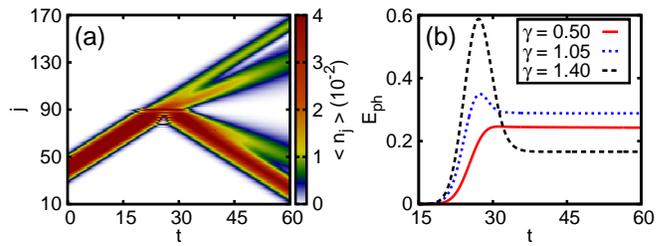}
\caption[short]{(Color online) (a) Electronic density $\langle n_j \rangle$ as a function of site $j$ and time $t$
for $L_H=1$, $\gamma=2.5$, and $\omega_0=1.65$. The position of the EP site is $j=90$.
(b) Phonon energy $E_{\text{ph}} = \omega_0 N_{\text{ph}}$ as a function of time for $L_H=1$, $\omega_0=1.5$, and several values of $\gamma$.}
\label{fig:dissipation}
\end{figure}

The second phenomenon is the dissipation of the electron energy due to inelastic scattering processes.  
 Figure~\ref{fig:dissipation}(a) shows that one pair of transmitted and reflected wave packets moves with the same
absolute velocity $v_0=2$ as the incident wave packet while a second pair moves with a lower velocity $v_1 \approx 1.1$~\cite{suppmat}.
This corresponds to an inelastic process  where a phonon is excited by the presence of the electron in the EP-coupled structure
and then left behind when the electron propagates away from this structure.
The final velocity $v_n=2t_0 \sin(k_n)$ can easily be determined from the equality of the asymptotic total energy
for $t\rightarrow \pm \infty$, 
$-2t_0 \cos(K) =  n \, \omega_0 - 2t_0 \cos(k_n)$, where $n$ is the number of excited phonons left behind.
Similar patterns have been observed recently in a 1D photonic wave guide coupled to a two-level scatterer,
see Fig.~3 in Ref.~\cite{sanc14}.
In Fig.~\ref{fig:dissipation}(b) we see that the phonon energy $E_{\text{ph}} = \omega_0 N_{\text{ph}}$,
which is zero initially, remains finite after the electron has left the EP-coupled structure (i.e., for $t \rightarrow \infty$).
This confirms that an irreversible energy transfer occurs from the electron to the phonon degrees of freedom (dissipation).

In summary, we have developed a TEBD algorithm with a dynamical optimization of local boson bases that allows us to simulate
the nonequilibrium dynamics of electron-phonon systems more efficiently. This opens the way for numerous
theoretical investigations of time-resolved spectral properties~\cite{baso11,orenstein12}, photoinduced phase 
transitions~\cite{nasu04,yone08}, and transport in low dimensions~\cite{galp07,osor08,zimb11}.
The overall performance depends
on the properties of the local density matrix out of equilibrium  and thus on the specific problem investigated.
The basic idea can easily be combined with other time-dependent MPS methods~\cite{whit04,dale04,schm04a}
or applied to other bosonic systems such as correlated photons~\cite{gruj12,caru13},
quantum baths~\cite{weiss2008,yao13}, and scalar fields~\cite{berg14} as well as to other systems with large local Hilbert
spaces such as high spin models~\cite{capp08,fuhr08}.

\begin{acknowledgments}
C.~B., F.~D., F.~H.-M., and E.~J. acknowledge support from the DFG (Deutsche Forschungsgemeinschaft) through
grants Nos.~JE~261/2-1 and HE~5242/3-1 in the Research Unit
\textit{Advanced Computational Methods for Strongly Correlated Quantum Systems} (FOR 1807).
L.~V. was supported by the Alexander-von-Humboldt foundation.
This work was also supported in part by National Science Foundation Grant No. PHYS-1066293 and the hospitality of the Aspen Center for Physics.
\end{acknowledgments}

\bibliographystyle{biblev1}
\bibliography{references}

\end{document}